# A Behavioral Analysis on the Reselection of Seed Nodes in Independent Cascade Based Influence Maximization

## Ali Vardasbi, Hesham Faili, Masoud Asadpour

School of Electrical and Computer Engineering, College of Engineering, University of Tehran, Tehran, Iran
{a.vardasbi, hfaili, asadpour}@ut.ac.ir

## ABSTRACT

Influence maximization serves as the main goal of a variety of social network activities such as viral marketing and campaign advertising. The independent cascade model for the influence spread assumes a one-time chance for each activated node to influence its neighbors. This reasonable assumption cannot be bypassed, since otherwise the influence probabilities of the nodes, modeled by the edge weights, would be altered. On the other hand, the manually activated seed set nodes can be *reselected* without violating the model parameters or assumptions. The *reselection* of a seed set node, simply means paying extra budget to a previously paid node in order for it to retry its influential skills on its uninfluenced neighbors. This view divides the influence maximization process into two cases: the *simple case* where the reselection of the nodes is not considered and the *reselection case*.

In this study we will analyze the behavior of real world networks on the difference between these two influence maximization cases. First we will show that the difference between the simple and the reselection cases constitutes a wide spectrum of networks ranging from the reselection-independent ones, where the reselection case has no noticeable advantage to the simple case, to the reselection-friendly ones, where the influence spread in the reselection case is twice the one in the simple case. Then we will correlate this dynamic to other influence maximization dynamics of the network. Finally, a significant entanglement between this dynamic and the network structure is shown and verified by the experiments. In other words, a series of conditions on the network structure is specified whose fulfilment is a sign for a reselection-friendly network. As a result of this entanglement, reselection-friendly networks can be spotted without performing the time consuming influence maximization algorithms.

## 1 INTRODUCTION

The focus on the influence maximization and influence propagation has been growing increasingly in the social network studies [1]–[5]. The fundamental question concerning the influence maximization problem is that what group of nodes, when selected as the initial influencers, can spread the desired influence to the highest extent possible [6]. The selection of a node as an initial influencer practically means spending a reasonable amount of budget such as money, time, reputation, etc. in order to activate it. An active node then tries to influence its neighbors and hopefully the cascade of influence would be triggered.

There are different theoretical models for the influence spread in a social network, amongst which the linear threshold (LT) and independent cascade (IC) models are the most used ones. In the LT model each node is considered to have a threshold and it is activated when the number of its active neighbors goes above that threshold [7], [8]. The IC model, on the other hand, deals with the influence probabilities of the links [9]. According to the IC model, each directed link $(u, v)$ is associated with a probability $p_{uv}$ that indicates the power of $u$ at influencing $v$. Once $u$ is influenced (either as an initial node or during the influence spread), it has a *onetime chance* to activate $v$ and is successful to do so with probability $p_{uv}$. During the influence spread process, giving the node $u$ a second chance to influence its neighbor $v$ will increase the influence probability from $p_{uv}$ to $1 - (1 - p_{uv})^2$ and the parameters of the IC model will be violated. However, when a previously influenced node is manually re-activated the scenario will be different. The difference between these two cases is more clarified in the following example.

Consider a social network for which the influence of individuals on their connections has been estimated from their activity. More specifically, in this example, influence has the form of clicking on the link that one has posted on the network. Furthermore, once a user has clicked on a posted link, his connections will be notified as if he has re-posted the link. Suppose that we have a web page and we desire to increase the number of our page views via advertising it on the mentioned social network. Our budget determines the number of initial users to whom we afford to introduce our page and ask them to post a link of it on the network. During the cascade of influence through the network, naturally a user will not re-post our link twice. Therefore, the connections of an active user will see his post once. But assume that we have paid one of our initial users double and asked him to post our link twice. Since the second chance has been given to him forcefully, the natural process of influence spread in the IC model has not been violated.

Furthermore, if the time interval between the two posts of the same user is selected appropriately, his influence power will be nearly doubled.

We will call the situation where a node is selected more than once, during the influence maximization process, the *reselection* of that node. It is worth noting that the reselection approach is quite common in the real world advertising. Usually, based on the budget of the company as well as the capacity of an advertising hub, the hub is paid more than once to popularize a specific product. Reselecting a hub to maximize the influence spread demonstrates the fact that when a node have a large number of important connections, a onetime attempt does not saturate its capacity and even if a fraction of its connections have been influenced at the first try, the hub's importance is still more than many other nodes in the network.

In this paper we study the dynamic of networks concerning the reselection of seed set nodes in an influence maximization process. We first evaluate the behavior of different real world networks against the reselection possibility of the seed set nodes. It is shown that different networks respond differently to this new feature. In some networks, there is hardly a duplicate in the first 50 nodes of the seed set. This means that, in the aforementioned networks, introducing a new node to the seed set usually has a better performance compared to reselecting a previous seed set node. On the other hand, in a number of other networks, only 10 to 20 percent of the first 50 seed set nodes are unique. These networks have a considerably higher influence spread when reselecting the seed set nodes is possible compared to the case where all the seed set nodes are required to be unique. We say that in the former networks the reselection feature has a low gain while in the latter networks its gain is high.

The main question of this study is about the structural cause of the above observation in social networks. To tackle this question, first it is shown that the gain of the reselection feature on the influence spread is correlated to another influence maximization dynamic, the *influence saturation*. Roughly speaking, the influence saturation measures the extent of degradation in the marginal influence spread during the expansion of the seed set nodes. Then, using the correlation between the influence saturation and the reselection gain, an entanglement between these dynamics and the network structure will be shown. The significance of this result is most understood for the large networks on which performing the influence maximization algorithms is time consuming.

The structure of the consequent sections is as follows. In Section 2 a brief overview of the influence maximization research is given. Section 3 is devoted to the definitions and parameters required for the following parts of the paper. We will prove in Section 4 that the influence maximization in the reselection case has the property of diminishing returns. In Section 5 we will discuss the saturation dynamic in the influence maximization and propose two parameters for measuring it. In order to be able to present our observations in the real world networks, we first explain our experimental setup in Section 6. After that, in Section 7, we first argue why the reselection gain is supposed to correlate to the saturation behavior and then confirm our discussion by the experimental results on 12 real world networks. In Section 8, we will define a new parameter for modeling the presence of the strong hubs in a network and show a correlation between this parameter and the reselection gain. Finally, we will conclude the paper and propose possible future works in Section 9.

## 2 RELATED WORKS

The formal definition of influence maximization is given in [6] as:

**Definition 1** (Influence Maximization) Given a graph G as a social network and a diffusion model for the influence; determine the set of influential targets of size at most k whose activation will cause the largest number of activated nodes in G.

Kempe et al. showed that the influence spread function is a submodular function and hence proposed a greedy $(1 - 1/e)$-approximation to the above problem. The high time complexity of the greedy algorithm commenced a new stream of research on the scalable influence maximization proposals. In this paper the CELF++ algorithm of [10] is referred to as the simple greedy algorithm. However, CELF++ and other speed ups such as [11], [12] did not scale acceptably for the networks of millions of nodes. As the social networks grow larger and larger, the need to scalable algorithms with promising performances becomes more realized. That is why a considerable number of scalable influence maximization algorithms have been published in recent years [13]–[18].

## 3 PARAMETERS AND DEFINITIONS

Considering the possibility of reselection at the influence maximization seed set nodes is equivalent to substitute *set* into its generalized concept *multiset*. A multiset is a collection of elements that can have multiple instances of elements [19]. The number of instances of an element in a multiset is called the element's *multiplicity*. For example in the multiset $\{a, a, a, b\}$ the elements $a$ and $b$ have multiplicity 3 and 1 respectively. A set is a special case of a multiset for which all the elements have multiplicity 1. Multisets are sometimes represented by elements of $\mathbb{Z}_+^m$, a vector of non-negative integers where $m$ is the size of the elements space and each field of the vector represents the multiplicity of the corresponding element.

Consequently, the *reselection possible influence maximization* is defined with the help of the multisets.

**Definition 2** (reselection possible influence maximization) Given a graph G as a social network and a diffusion model for the influence; determine the seed multiset of influential targets of size at most k whose activation will cause the largest number of activated nodes in G. Each node of the seed multiset with multiplicity $m$ has a $m$ times chance at influencing its neighbors.

One may argue that the reselection of a seed node has less influence compared to its selection as the first time. To address this issue, we define a more general setting that models the possible fading effect caused by reselection.

**Definition 3** (reselection possible influence maximization with fading) Given a graph G as a social network, a diffusion model for the influence and a fading parameter $0 \leq \alpha \leq 1$; determine the seed multiset of influential targets of size at most k whose activation will cause the largest number of activated nodes in G. Each node of the seed multiset with multiplicity $m$ has a $m$ times chance at influencing its neighbors; but its influence at the $\omega^{\text{th}}$ chance is faded by a factor of $\alpha^{\omega-1}$. The extreme cases where $\alpha=0$ or $\alpha=1$ respectively correspond to the simple influence maximization case (Definition 1) and the reselection possible influence maximization without fading (Definition 2).

Submodular functions play an important role in influence maximization as well as a great number of computer science optimization problems. A submodular function is mostly known by the diminishing return property.

**Definition 4** (Submodular function) A set function $f: 2^V \rightarrow \mathbb{R}$ is submodular if for every $A \subseteq B \subseteq V$ and $e \in V \setminus B$ it holds that

$$f(A \cup \{e\}) - f(A) \geq f(B \cup \{e\}) - f(B). \quad (1)$$

This property is known as the property of diminishing returns. In [20] the property of diminishing returns is nicely extended to multisets as follows.

**Definition 5** (Diminishing returns on multisets) A function $f: \mathbb{Z}_+^m \rightarrow \mathbb{R}$ has the diminishing returns property if for every $x, y \in \mathbb{Z}_+^m$ such that $x \leq_{\mathbb{Z}_+^m} y$ and any unit base vector $e_i = (0, \cdots, 0, 1, 0, \cdots, 0) \in \mathbb{Z}_+^m$, it holds that

$$f(x + e_i) - f(x) \geq f(y + e_i) - f(y). \quad (2)$$

Through the rest of this paper, the influence spread function of a set $S$ and a multiset $M$ on a network $G$ is shown by $\sigma_G(S)$ and $\sigma_G^m(M)$, respectively. The superscript $m$ on the latter function denotes the multiset domain of the function. To compute the spread of $M$, each node of the multiset is given as many chances as its multiplicity within $M$.

Finally, we define the *reselection gain* (RG) to be the ratio of the influence spread in the reselection case to the simple case. Formally, for a given graph $G$ and seed size $k$, the reselection gain is defined to be:

$$RG_G(k) = \frac{\max\limits_{|M|=k} \sigma_G^m(M)}{\max\limits_{|S|=k} \sigma_G(S)} \quad (3)$$

## 4 SUBMODULARITY OF RESELECTION

In this section it is proved that the spread function on the multisets of nodes has the diminishing returns property. As a result the greedy algorithm is the best achieved provable approximation for the reselection possible influence maximization as well.

Intuitively, a node at its $k^{\text{th}}$ chance for influencing its neighbors cannot influence more neighbors (on average) than it had influenced at its $(k-1)^{\text{th}}$ chance. So for a specific node we expect to see a concave function of the node's spread in terms of its number of chances. This concave function when combined with the diminishing returns property of the spread function on the sets, will result to a spread function on the multisets with diminishing returns. In what follows, this intuition will be proven. In this proof we will use the result from [6] which states that the spread function on the sets in the IC model is a monotone submodular function.

**Theorem 1.** The spread function $\sigma_G^m(M)$ on multisets of nodes has the diminishing returns property as defined in Definition 5.

*Proof.* For the given graph $G(V_G, E_G)$ construct a new graph $H(V_H, E_H)$ in which for every node $v \in V$ there are $n$ copies $v_1, v_2, \cdots, v_n$ (for our purposes it suffices to let $n = |V|$; but it can be set as large as required). Connect each copy $v_i$ to all the out-neighbors of the original $v$ with an influence probability multiplied by $\alpha^i$. Since the copied nodes have no incoming link, they do not participate in the influence spread. Their only contribution is when they are activated manually. It is easy to see that reselecting a node $v$ for $m$ times (giving it $m+1$ times chance to influence others) has the same impact on the neighboring nodes as selecting $m$ copies $v_1, v_2, \cdots, v_m$ of $v$ on $H$. Since these $m$ copies are also computed in the spread function of $\{v, v_1, \cdots, v_m\}$, the following equality holds:

$$\sigma_G^m\left(\left\{\underbrace{v, v, \cdots, v}_{m+1}\right\}\right) = \sigma_H(\{v, v_1, v_2, \cdots, v_m\}) - m \quad (4)$$

For a multiset $M$ of $V_G$, we define the $M_{set} \subseteq V_H$ to be the underlying set of $M$. $M_{set}$ is simply obtained by substituting all the repetitions of a node in $M$ by its copies in $V_H$. For example, the underlying set of $\{v, v, v, u, u\}$ is $\{v, v_1, v_2, u, u_1\}$. The number of unique elements of a multiset $M$ is shown by $u(M)$. From (4) it is straightforward that:

$$\sigma_G^m(M) = \sigma_H(M_{set}) - |M_{set}| + u(M) \quad (5)$$

Now we show that the **Definition 5** holds for the $\sigma_G^m$ function. Consider two multisets $X$ and $Y$ of $V_G$ and a node $v$. The multisets are such that $X_{set} \subseteq Y_{set}$. Suppose that the multiplicity of $v$ in $X$ and $Y$ is shown by $j_x$ and $j_y$ respectively. The following relations hold:

$$\sigma_H(X_{set} \cup v_{j_x}) - \sigma_H(X_{set}) \geq \sigma_H\left(Y_{set} \cup v_{j_y}\right) - \sigma_H(Y_{set}) \quad (6)$$

$$\left|X_{set} \cup v_{j_x}\right| - |X_{set}| = \left|Y_{set} \cup v_{j_y}\right| - |Y_{set}| = 1 \quad (7)$$

Concerning the selection of $v$, three cases may happen: (I) $v \in X$, (II) $v \notin Y$ or (III) $v \in Y \& v \notin X$. It is not hard to see that the following inequality holds in all of these three cases:

$$u(X_{set} \cup v_{j_x}) - u(X_{set}) \geq u\left(Y_{set} \cup v_{j_y}\right) - u(Y_{set}). \quad (8)$$

Finally, combining (6), (7) and (8) proves the diminishing returns property for the $\sigma_G^m$ function. ∎

Theorem 1 together with the monotonicity of $\sigma_G^m$ shows that the greedy algorithm in the reselection possible influence maximization case performs as good as in the simple case. The only difference in implementing the greedy algorithm is that unlike the simple influence maximization case, the selected node at each iteration would not be dismissed in the reselection case.

## 5 INFLUENCE SATURATION

Suppose that for each $k$ the maximum influence spread on graph $G$ caused by activating $k$ nodes of $G$ is shown by $\tau_G(k)$. The submodularity of the spread function implies that $\tau(k)$ is a concave function of $k$. As such, for every graph $G$ there is a saturation threshold $k_G^*$ after which the positive slope of the $\tau_G(k)$ function will be insignificant; i.e. the graph *saturates* by the influential seed set nodes of size $k_G^*$.

Observations on the behavior of the $\tau_G(k)$ function for real world networks $G$ reveals an interesting saturation dynamic. For a number of networks the saturation threshold is 1. In other words, the influence spread of the most influential node is such that the marginal gain of the next seed set nodes becomes negligible. We call this behavior as the *sharp saturation*. Figure 1 shows two networks with different saturation behaviors. The y-axis of these plots is the $\tau_G(k)$ normalized by the node size of graph $|G|$ for simplicity of comparison. As can be seen in the figure, the Slashdot (the networks will be introduced in Section 6) graph has a sharp saturation.

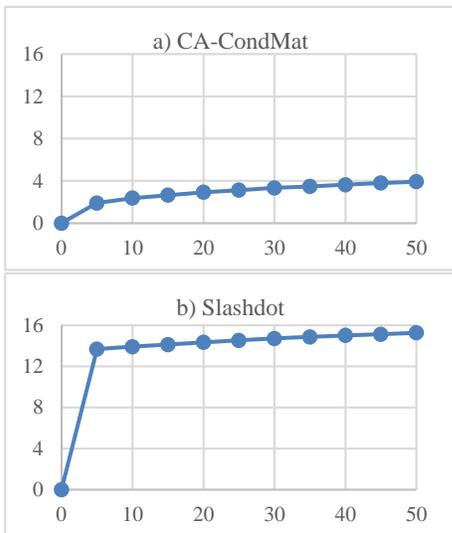

**Figure 1 The plot of $\tau_G(k)/|G|$ versus $k$. The CA-CondMat network (a) has a smooth curve; while the Slashdot network (b) has a sharp saturation**

We define the *influence saturation* (IS) parameter to entail the saturation dynamics of different networks. For fixed values of $k_{min}$ and $k_{max}$ (in this paper we set $k_{min} = 5$ and $k_{max} = 50$) suppose that the linear approximation of the function $\tau_G(k)$ inside the $[k_{min}, k_{max}]$ is as follows:

$$\hat{\tau}_G(k) = \sigma_1(G) \cdot k + \sigma_0(G) \quad ; \quad (k_{min} \leq k \leq k_{max}). \quad (9)$$

This approximation tells us that, observing from the linear world, the first influential node has influenced $\sigma_1 + \sigma_0$ nodes while the next influential seed set nodes has influenced only $\sigma_1$ nodes on average. A very high $\sigma_1 + \sigma_0$ to $\sigma_1$ ratio indicates a sharp saturation while a low ratio shows a smooth saturation. Therefore, it is natural to define the influence saturation parameter as follows:

$$IS(G) = \frac{\sigma_1(G) + \sigma_0(G)}{\sigma_1(G)} \quad (10)$$

where $\sigma_1$ and $\sigma_0$ are obtained from (9).

The IS and RG parameters are expected to be high in a star like network. By the star like network we mean a network whose nodes can be decomposed into two components:

- Core nodes: a small set of nodes which are connected to a considerable fraction of network. These nodes are highly influential.

- Loosely connected nodes: a large number of nodes which are weekly connected to each other but strongly connected to the core nodes.

In a star like network, selecting one of the core nodes will spread the influence to a large section of the network and causes a sharp saturation. On the other hand, reselecting the core nodes instead of non-core nodes is likely to increase the influence spread which means a high reselection gain.

In our experiments we will test the hypothesis that "are all the networks with a high RG, star like?" To test this hypothesis we will use the above argument about the entanglement of IS and RG in the star like networks.

## 6 EXPERIMENTAL SETUP

The experiments of this paper are conducted on the real world networks obtained from [22]. The node and edge sizes of the networks range from 4k to 317k and 28k to 2M respectively. All the networks in this paper are directed. In the cases where the original network was undirected, we have considered two directed edges for each undirected edge, making the edge size of the network twice its original. The networks are described bellow:

- **Facebook**: The Facebook dataset consists of friend lists from Facebook. The data is collected from survey participants [23]. In our experiments we only used the graph of friendship.

- **Wiki-Vote**: The network contains all the Wikipedia adminship voting data until January 2008. The nodes represent wikipedia users and a directed edge from node $i$ to node $j$ indicates that user $i$ has voted for the adminship of user $j$ [24], [25].

- **Email-Enron**: This dataset contains the email communications of Enron. The nodes represent the Enron email addresses and an undirected link between $i$ and $j$ indicates that either of them has sent an email to the other [26], [27].

- **Epinions**: This graph is a who-trusts-whom online social network of a general consumer review site Epinions.com [28].

- **Slashdot**: Slashdot is a technology-related news website known for its specific user community. The network contains friend/foe links between the users of Slashdot [27].

- **DBLP**: The DBLP computer science bibliography provides a comprehensive list of research papers in computer science. This graph is a co-authorship network where two authors are connected if they publish at least one paper together [29].

- **CA-GrQc, CA-HepTh, CA-HepPh, CA-Astro, CA-CondMat**: These graphs are the collaboration network from the e-print arXiv and covers scientific collaborations between authors papers submitted to General Relativity and Quantum Cosmology category, High Energy Physics Theory, High Energy Physics Phenomenology, Astro Physics and Condense Matter categories, respectively [30].

- **Cit-HepPh**: The citation graph from the e-print arXiv that covers all the citations of High Energy Physics Phenomenology papers. A directed link from paper $i$ to $j$ indicates that paper $i$ cites paper $j$ [31], [32].

The network statistics are shown in Table 1.

As is common in the influence maximization research on the IC model, for the edge weights we use the following two models:

- **Weighted Cascade (WC) model**: In the WC model, the influence probability of each edge is assigned to $P_{uv} = 1/d_v$, where $d_v$ is the in-degree of $v$ [6].

**Table 1  Network statistics**

| Network | #nodes | #edges |
|---------|--------|--------|
| Facebook | 4,039 | 176,468 |
| Wiki-Vote | 7,115 | 103,689 |
| Email-Enron | 36,692 | 367,662 |
| Epinions | 75,879 | 508,837 |
| Slashdot | 77,360 | 905,468 |
| DBLP | 317,080 | 2,099,732 |
| CA-GrQc | 5,242 | 28,980 |
| CA-HepTh | 9,877 | 51,971 |
| CA-HepPh | 12,008 | 237,010 |
| CA-AstroPh | 18,772 | 396,160 |
| CA-CondMat | 23,133 | 186,936 |
| Cit-HepPh | 34,546 | 421,578 |

- **Trivalency (TR) model:** This model assigns a randomly selected probability from {0.1, 0.01, 0.001} to each directed link [12].

We set the seed set/multiset size of the influence maximization algorithm to be 50 nodes. Since the simple greedy algorithm requires a prohibiting long time on large networks, in cases of Epinions, Slashdot and DBLP networks (and only on these networks) we use the IMM algorithm [17] for the simple influence maximization case rather than the greedy algorithm. The IMM algorithm has been shown, theoretically and practically, that performs nearly as good as the greedy algorithm.

For the reselection case on the aforementioned large networks (i.e. Epinions, Slashdot and DBLP), we run the greedy algorithm for multiset selection on the 50 nodes obtained from the IMM algorithm. The intuition behind this is that the influential multiset in the reselection case is obtained by reselecting some of the nodes from the influential set in the simple case and removing some other nodes as a result of this reselection. This intuition has been confirmed on the small networks for which the greedy algorithm is feasible.

## 6.1  RESELECTION IMPACT

In this section, before studying the relation between the previously defined parameters, we show the impact of the reselection with varying fading values on different networks. Based on their influence spread behavior in response to the reselection possibility, we categorize the network into the three following cases:

- **Reselection friendly networks**: When the reselection gain in a network without any fading ($\alpha$=1) is more than 1.5 we call it a reselection friendly network. In these networks the possibility of the reselecting the nodes increases the influence spread more than 50% compared to the simple case. A simple example of a reselection friendly network is a star graph consisting of a core node and a number of pairwise disjoint nodes connected only to the core node. Obviously, reselecting the core node multiple of times has an outstanding gain compared to the simple case where the core node can only be selected once.

- **Reselection aware networks:** In the absent of fading ($\alpha$=1) when the reselection gain of a network lies between 1.05 and 1.5, the network is called to be reselection aware. The impact of reselection on these networks is not as impressive as the previous case; but it is noticeable.

- **Reselection free networks:** These networks have a reselection gain less than 1.05. In the reselection free networks the multiset obtained by solving the reselection possible influence maximization hardly differs from the solution of the simple influence maximization case. A good example of such networks is a clique with uniform influence probabilities. In a fully connected network all the nodes share the same set of neighbors and reselection of a node has almost the same influence as selecting a new node.

Figure 2 plots the changes of the reselection gain in terms of the fading parameter α when the influence probabilities are derived from the WC model. As can be seen in this figure, Facebook and Wiki-Vote networks are reselection friendly networks (Figure 2-a), CA-AstroPh, CA-CondMat, CA-HepPh and Email-Enron networks are reselection aware (Figure 2-b) and Cit-HepTh, CA-GrQc and CA-HepTh networks are reselection free (Figure 2-c). It is interesting to note that the reselection gain in the reselection friendly networks, even with a fading value as low as α=0.6 is still non-negligible.

Surprisingly, no one of the tested networks in the TR model are reselection friendly. Figure 3 illustrates the change of reselection gain in terms of fading value α for the TR model. A comparison between Figure 2 and Figure 3 shows that the behavior of the networks is totally dependent to the influence probability model. More specifically, the following scenarios are observed in these figures:

- Facebook and Wiki-Vote networks are reselection friendly in WC model but reselection aware in TR model;
- CA-HepTh and CA-GrQc networks are reselection aware in TR model but reselection free in WC model;
- CA-HepPh and CA-AstroPh networks are reselection aware in WC model but reselection free in TR model.

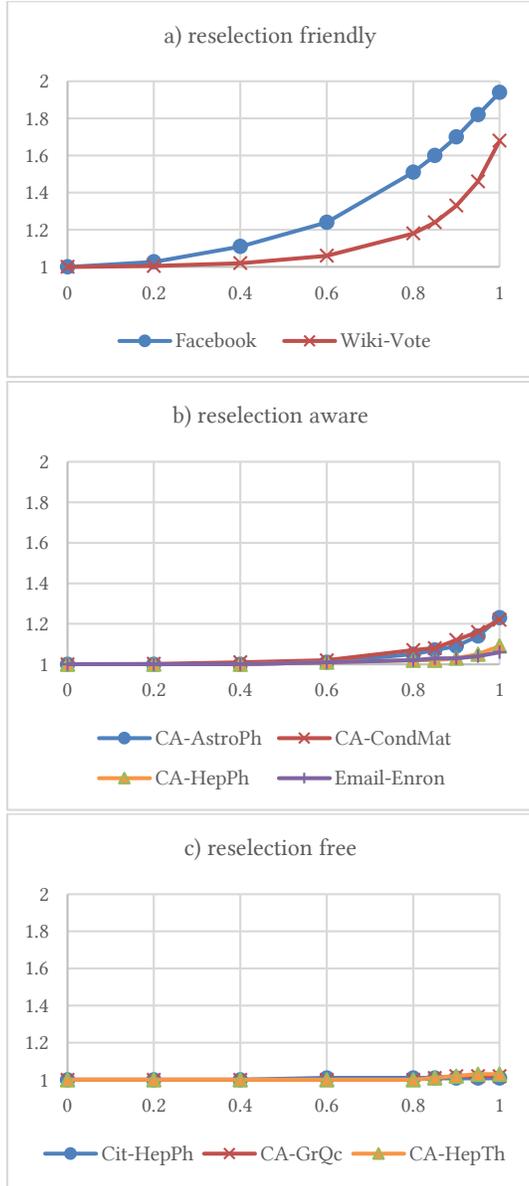

**Figure 2 Reselection gain of different networks with WC model in terms of the fading value (α)**

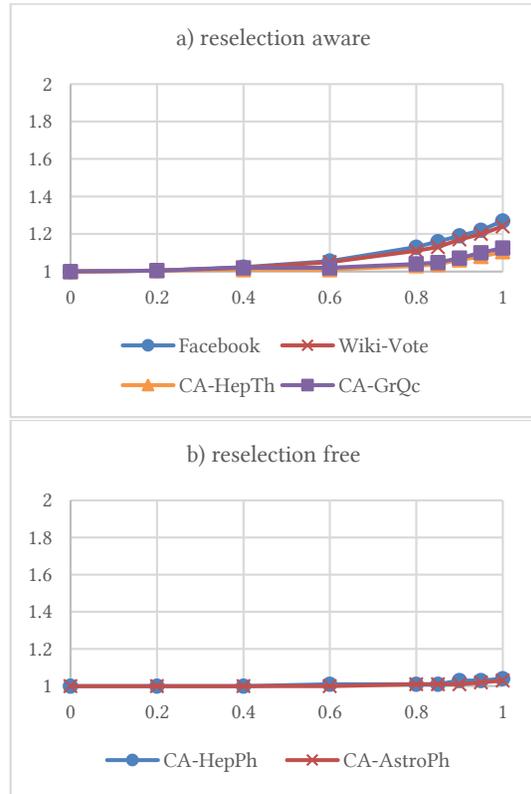

**Figure 3 Reselection gain of different networks with TR model in terms of the fading value (α)**

## 6.2 CORRELATION BETWEEN *RG* AND *IS*

In our first experiment, we have studied the correlation between the reselection gain (RG) and influence saturation (IS) on the real world networks mentioned in Section 6.

When the IS parameter is high in a network, it means that the influence spread of the first seed node is considerably higher than the marginal influence spread of the next seed nodes. The structural interpretation of this dynamic is that the network contains a dense core with two important properties: (I) the density of the core is such that an influential node within the core can influence a great portion of the core; and (II) the strength of the connections from the in-core nodes to the out-core nodes is such that the activated core nodes can influence a great number of outer nodes.

On the other hand, a high RG ratio suggests the presence of strong hubs in the network. In the context of influence maximization, a hub usually has two properties: (I) it has a significant number of strong connections; and (II) its connections, when activated, can in turn influence a considerable number of nodes.

Even though the above situations for the cause of a high IS and a high RG does not necessarily translate to each other, they have a positive correlation in real world networks with the WC model. On the contrary, when the TR model is considered, no meaningful correlation is observed between the IS and RG. Consequently, in this section we only report the results for the WC model and leave the question of "why the RG and IS parameters are highly correlated in WC model but uncorrelated in TR model?" as an open problem.

Figure 4 shows the influence spread in the simple and reselection cases in a number of our tested networks in the WC model. It also contains the linear approximation of the $\tau(k)$ function. As can be seen in this figure, networks such as

Facebook and Slashdot with a sharp saturation have a high RG ratio, while Email-Enron and DBLP with a smooth saturation have a RG ratio near the unity.

Table 2 shows *IS* and *RG* parameters of the networks. Using the values presented in Table 2, the RG ratio has a significant correlation of about 0.78 to the IS parameter.

Finally, our earlier hypothesis of high RG networks being star like is found to be model dependent. For the WC model, we have observed that high RG networks are likely to have a high IS and therefore star like. However, for the TR model, there are high RG networks with low IS and vice versa. This means that the star like networks are not the dominant contributors for the high RG networks in the TR model.

## 6.3 NETWORK STRUCTURE

As was discussed in previous section, a high RG is supposed to happen when strong hubs exist in the network. The presence of strong hubs can be detected by comparing the influence spread of nodes. As for the hubs, usually there exists a gap between their influence spread and the spread due to the ordinary nodes.

For each network, we first sort the nodes based on their influence spread and then plot the highest influential nodes. The presence of hubs is identified by a fast degradation in this plot. As such, we are left with two types of networks: the ones whose RG is correctly predicted by their hubs (Figure 5) and the ones for which our prediction is wrong (Figure 6).

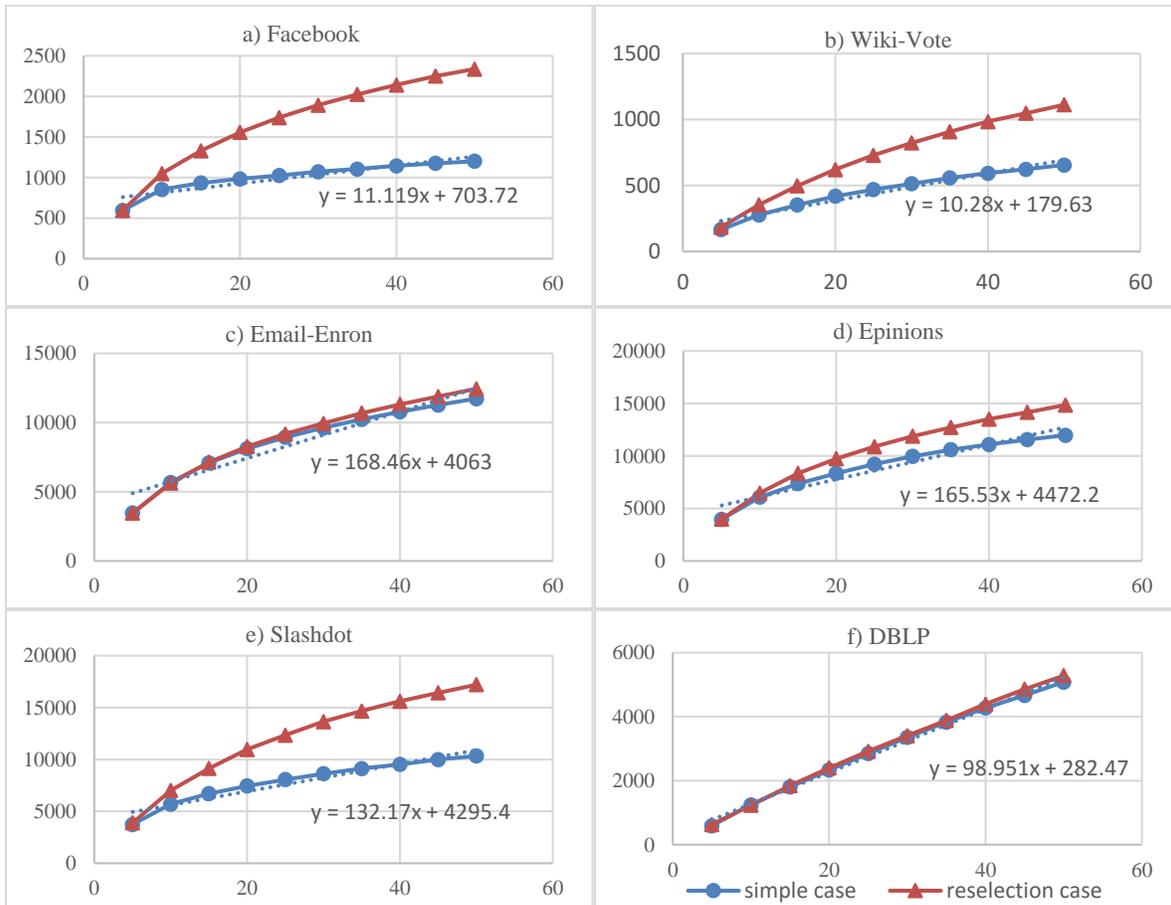

**Figure 4 Influence spread and its linear approximation in the simple case versus the reselection case**

**Table 2   The IS and RG parameters of the networks with WC model**

| Network | IS | RG |
|---|---|---|
| Facebook | 64.29 | 1.94 |
| CA-GrQc | 7.76 | 1.02 |
| Wiki-Vote | 18.47 | 1.70 |
| CA-HepTh | 7.83 | 1.05 |
| CA-HepPh | 13.40 | 1.07 |
| CA-AstroPh | 17.18 | 1.21 |
| CA-CondMat | 12.55 | 1.22 |
| Cit-HepPh | 15.90 | 1 |
| Email-Enron | 25.12 | 1.06 |
| Epinions | 28.02 | 1.24 |
| Slashdot | 33.50 | 1.67 |
| DBLP | 3.85 | 1.04 |

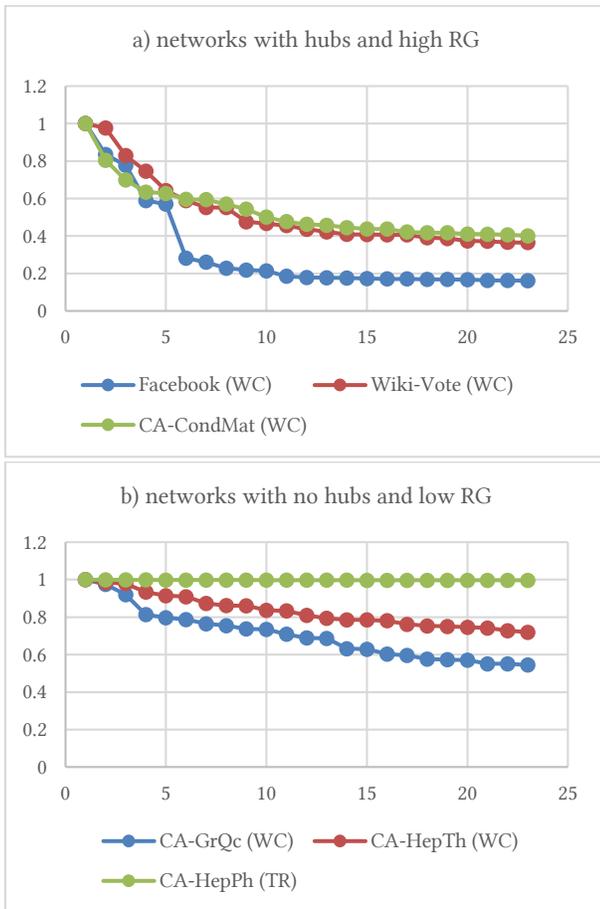

**Figure 5 Example networks where our hub detection correctly predicts a high/low reselection gain**

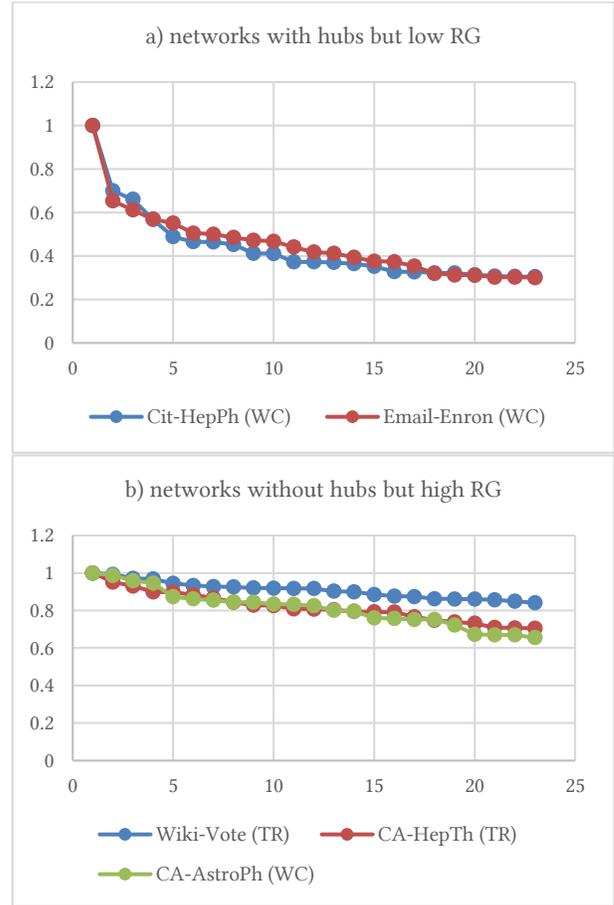

**Figure 6 Example networks where our reselection gain prediction based on hubs are wrong**

In order to give a quantitative measure for this method, we suggest to consider the ratio between the spread of the highest influential node to the $k^{th}$ influential node. Hereafter we call this parameter as $HR_k$ which stands for the Hub Ratio with distance k. This value measures the amount of degradation in the influence plot and the k acts as a smoothing parameter.

Figure 7 depicts the sensitivity of $HR_k$ to the smoothing parameter k. The correlation is taken on the 24 values obtained from considering all 12 networks in both WC and TR models. As can be seen in this figure, the correlation has a stable value around 70% for different smoothing values. Unlike the IS parameter, the HR parameter is suitable in both WC and TR models.

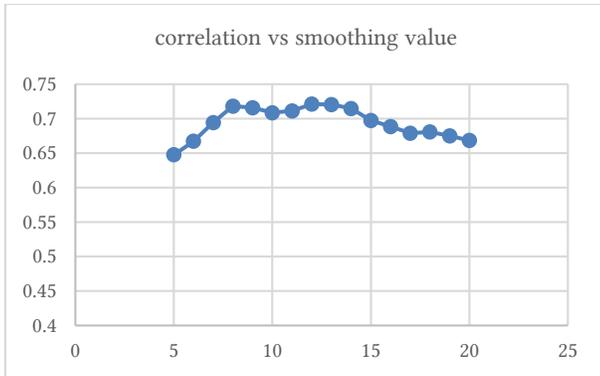

**Figure 7 Correlation between the $HR_k$ of all the 12 tested networks in both WC and TR models in terms of k**

## 7 CONCLUSION

In this paper we have seen that considering the possibility of node reselection in the influence maximization, or equivalently targeting multiset of seeds instead of set of seeds can improve the influence spread in a number of networks. However, there is no guarantee that using the reselection possible influence maximization have a considerable gain over the simple case. Our experiments have shown that the reselection gain can vary from 1 to 1.9 in different real world networks.

We have correlated the different reselection gains of different networks to another influence maximization dynamic, called the influence saturation. We have shown experimentally that there is a 0.8 correlation between the reselection gain and the influence saturation in our tested networks in the WC model.

Finally, in a search for a measurement for the presence of strong hubs in a network, we have introduced the hub ratio parameter and shown a correlation of about 0.7 between the reselection gain and the hub ratio in both WC and TR models. We think that there are still room for analyzing the reselection gain difference in different networks. Finding a stronger entanglement between this dynamic and the network structure enables us to distinguish the networks with high reselection gain from the networks with no reselection gain and choose our advertising strategies accordingly.